\documentclass[10pt,letterpaper,compsoc,conference]{IEEEtran}

\usepackage{cite}
\usepackage{amsmath,amssymb,amsfonts}
\usepackage{algorithmic}
\usepackage{graphicx}
\usepackage[dvipsnames]{xcolor}
\usepackage[final]{microtype}
\usepackage[italic]{mathastext}
\usepackage{libertine}
\usepackage[T1]{fontenc}
\usepackage{textcomp}
\usepackage[varqu,varl]{zi4}
\usepackage[all]{nowidow}
\usepackage{fancyhdr}

\usepackage{orcidlink}
\usepackage{parskip} 
\usepackage{array} 
\usepackage{enumitem}
\setlist[itemize]{itemsep=0.1em, topsep=0.1em} 
\usepackage{balance}
\usepackage{todonotes}  

\fancypagestyle{firstpage}{
  \fancyhf{}
  
  \fancyhead[C]{\vspace{10pt}\normalsize{Preprint working draft.} \\\vspace{-25pt}} 
  \fancyfoot[C]{\thepage}
}

\begin{document}

\title{Characterizing the Variance Envelope: A Multi-Dimensional Analysis of Spectre Telemetry Across Architectures and Workloads}

\author{
    \IEEEauthorblockN{Jaya Keshava Chandra Kotha\orcidlink{0009-0004-0865-9990}}
    \IEEEauthorblockA{\textit{Department of Electrical Engineering and Computer Science} \\
    \textit{University of California, Irvine}\\
    Student member, IEEE}
    \and
    \IEEEauthorblockN{Jean-Luc Gaudiot\orcidlink{0000-0001-9164-8731}}
    \IEEEauthorblockA{\textit{Department of Electrical Engineering and Computer Science} \\
    \textit{University of California, Irvine}\\
    Life Fellow, IEEE}
}

\maketitle
\thispagestyle{firstpage}
\pagestyle{plain}


\begin{abstract}

%
%
%

Hardware attacks like Spectre exploit built-in processor vulnerabilities, leaving anomalous footprints in Hardware Performance Counter (HPC) metrics. While machine learning can detect these footprints in controlled settings, static models fail in the real world when confronted with background system noise, diverse attack variants, and adversarial traffic pacing. To close this gap, this paper characterizes the ``variance envelope''---the full range of how attack signatures shift---across Intel, ARM, and AMD architectures. We evaluate an extensive experimental matrix encompassing three attack variants, four pacing modes, and four background-noise conditions. Our analysis proves that HPC signatures are highly fragile and easily warped by their execution environment. Furthermore, we expose a critical microarchitectural bottleneck: the persistent, hardware-level failure of Prime+Probe attacks on the AMD Jaguar. Ultimately, this comprehensive characterization demonstrates the fundamental limits of static detection thresholds. By proving that signatures are deeply intertwined with underlying execution environments, this study establishes the empirical foundation required for future work in advanced adaptive modeling and hybrid detection architectures.
\end{abstract}
\section{Introduction}
\label{sec:introduction}
%

\noindent Hardware security operates as a continuous arms race. As vulnerabilities like Spectre~\cite{koch2020spectre} emerge, researchers increasingly deploy Machine Learning (ML) models that monitor Hardware Performance Counters (HPCs) to detect the anomalous telemetry generated by transient execution attacks.

While prior works~\cite{kotha2025datadriven,li2022detecting,larson2021realtime,zhang2018cloudradar,zhou2021hardware} obtain near-perfect detection accuracy in isolated environments, these static ML detectors suffer from an illusion of portability and frequently fail in real-world systems. This fragility stems from evaluating a single attack mechanism on a single microarchitecture under idealized, continuous execution. In reality, attacker behavior, background system noise, and the underlying hardware interact to severely distort HPC telemetry distributions.

The detection challenge is further compounded by the diversity of side-channel leakage mechanisms. Most detectors focus on \textit{Flush+Reload} techniques, which rely on the direct timing of shared memory lines, or \textit{Prime+Probe} techniques, which infer data by measuring set-based disturbance across the cache hierarchy. Because these two primitives interact with the cache controller in fundamentally different ways, a detector trained exclusively on one often remains blind to the microarchitectural footprint of the other.

Recent Systematization of Knowledge (SoK) studies confirm that current detectors fail under realistic threat models because they are trained and evaluated in artificially constrained environments~\cite{kosasih2024sok}. Specifically, the field lacks a well-defined characterization of how Spectre HPC signatures manifest across varying workloads and hardware configurations. This absence of a stable threat signature means current models cannot reliably distinguish between malicious activity and stochastic benign fluctuations.

Building upon early foundational work that establishes both the feasibility of ML detection~\cite{li2019detecting} and the threat of adversarial pacing~\cite{li2022detecting}, we argue that a model cannot adapt to a new domain until its \textit{variance envelope}---the full range of how attack signatures shift under varying conditions---is thoroughly mapped. To answer the community’s call for realistic threat modeling, this paper presents a comprehensive empirical characterization of Spectre HPC signatures. Our core contributions are:

\begin{itemize}[itemsep=0.6em]
    \item \textbf{Multi-Variant and Multi-Architecture Profiling:} We evaluate telemetry shifts across three attack mechanisms (Spectre V1, V2) and two primary leakage channels (Flush+Reload and Prime+Probe) across four devices spanning three architectures (Intel, ARM, and AMD)\footnote{\textit{A Note on Hardware Nomenclature:} The specific devices evaluated in this study are treated as individual microarchitectural instances. They are not intended to represent the entirety of their respective Instruction Set Architectures (ISAs) or broader product families. We utilize these labels throughout the paper strictly for clarity and usability of reference to our specific testbeds.}
    \item \textbf{System Interference and Background Noise:} We measure signal degradation under diverse system loads (Idle, CPU, Memory, Mixed) to identify which HPC metrics are resilient and which collapse under environmental pressure.
    \item \textbf{Adversarial Pacing and Traffic Shaping:} We quantify how attacker evasion tactics (such as rate-limiting, burst, and batch execution) alter the stability of HPC telemetry.
    \item \textbf{Hardware-Intrinsic Attack Mitigation:} We document how specific microarchitectural constraints, such as the persistent failure of the Prime+Probe exploit on the AMD Jaguar architecture, act as inherent defenses. By proving that specific topological cache constraints—such as strict bank-routing structures—actively neutralize the attack, we demonstrate why cross-platform signature uniformity is physically impossible.
\end{itemize}

This paper provides a novel workload characterization of speculative attack signatures throughout different microarchitectural domains. We demonstrate that detection resiliency is inherently constrained by a telemetry-workload gap, where architectural bottlenecks and environmental noise cause a catastrophic collapse in signal separability.

The remainder of this paper is structured as follows. Section~\ref{sec:background} outlines the observability channels and Section ~\ref{sec:related_work} dives into the evolution of HPC detection. Section~\ref{sec:methodology} details our experimental matrix. Sections~\ref{sec:characterization} through~\ref{sec:bottlenecks} analyze the domain shifts caused by architecture, noise, and pacing. Finally, Section~\ref{sec:conclusion}discusses the broader implications of these microarchitectural bottlenecks and outlines our avenues for future work, specifically the necessity for advanced adaptive modeling and hybrid detection architectures.
\section{Background}
\label{sec:background}

\noindent
To understand why static detection models fail across different computing environments, we must first establish how transient execution attacks generate observable telemetry. This section details the distinct phases of a speculative attack—from the initial branch misprediction to cache-based exfiltration—and explains how underlying architectural differences dictate the resulting Hardware Performance Counter (HPC) signatures.

\subsection{Transient Execution and Leakage Channels}
\noindent
While all Spectre-family attacks exploit speculative execution, their demands on the microarchitecture diverge significantly based on the mechanism and the channel.

\subsubsection{Attack Mechanisms} 
\begin{figure}[ht]
    \centering
    \includegraphics[width=1.0\linewidth]{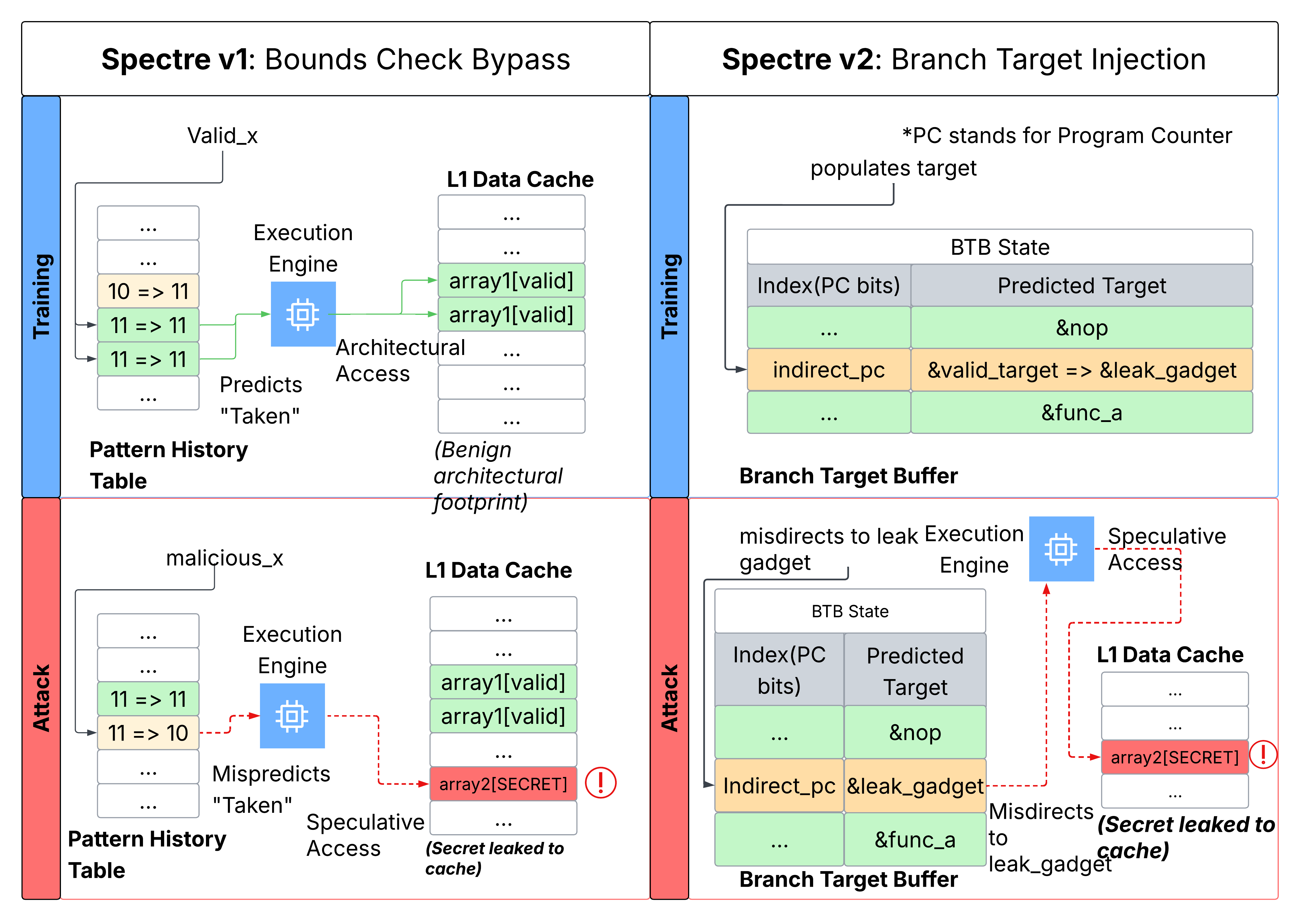}
    \caption{Microarchitectural hardware schematic contrasting Spectre variants. (a) Spectre V1 exploits the conditional Pattern History Table (PHT). (b) Spectre V2 poisons the indirect Branch Target Buffer (BTB).}
    \label{fig:spectre_microarc}
\end{figure}

Spectre V1 (Bounds Check Bypass)~\cite{koch2020spectre, canella2019systematic, mcilroy2019spectre} coerces the processor into speculatively executing instructions past a mispredicted bounds check. As illustrated in Figure~\ref{fig:spectre_microarc}(a), the attacker trains the Pattern History Table (PHT) to cause speculative secret access, which manifests in HPC telemetry as anomalous spikes in conditional branch mispredictions. Conversely, Spectre V2 (Branch Target Injection)~\cite{koch2020spectre, koruyeh2018spectre} poisons the indirect branch predictors, such as the Branch Target Buffer (BTB). As detailed in Figure~\ref{fig:spectre_microarc}(b), this shifts the microarchitectural footprint toward indirect branch speculation metrics.

\subsubsection{Observability Channels} 
\begin{figure}[ht]
    \centering
    \includegraphics[width=1.0\linewidth]{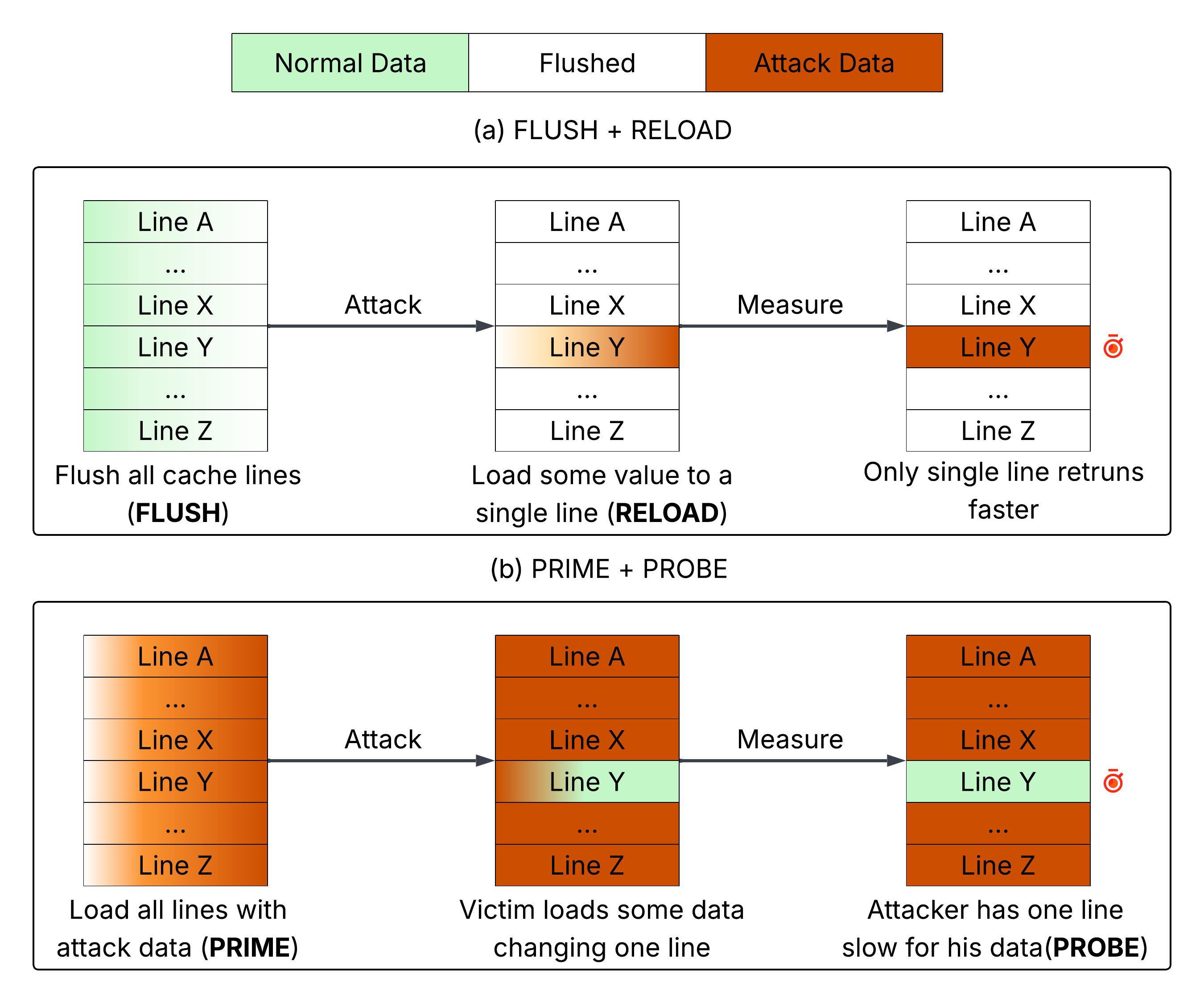}
    \caption{Comparison of cache side-channel leakage mechanisms. (a) Flush+Reload relies on shared memory lines. (b) Prime+Probe measures set-based disturbance.}
    \label{fig:cache_channels}
\end{figure}

Once sensitive data is speculatively accessed, it must be transmitted via a covert channel. The \textit{Flush+Reload} channel~\cite{yarom2014flush, gruss2016flush} relies on direct timing differences of shared memory lines. As shown in Figure~\ref{fig:cache_channels}(a), it generates a highly precise, surgical signature of cache hits and misses. In contrast, \textit{Prime+Probe}~\cite{percival2005cache, liu2015last, tromer2010efficient} infers data by measuring set-based disturbance across the cache hierarchy. As depicted in Figure~\ref{fig:cache_channels}(b), the reliance on eviction set construction and broader cache contention significantly changes the resulting telemetry profile.

\subsection{HPC Semantics and Cache Hierarchy}
\noindent
HPCs are microarchitectural features; their semantics are strictly hardware-dependent. The most critical divergence for side-channel observability lies in the \textbf{cache inclusion policy}~\cite{yan2017secure}. In inclusive hierarchies (predominant in many Intel processors), the Last Level Cache (LLC) acts as a directory for all lower-level lines. Saturating an LLC set triggers an automatic ``back-invalidation'' that forces data out of the L1, creating a high-magnitude, detectable HPC signature. In contrast, alternative topologies (such as the banked L2 structure of the AMD Jaguar) enforce strict physical memory routing, which severely misaligns standard, unified eviction sets and prevents reliable L1 back-invalidation~\cite{li2022detecting}.

\section{Related Work}
\label{sec:related_work}

\noindent
The use of hardware performance counters for security evolves through three distinct research epochs, moving from general malware classification to the current recognition of signature fragility.

\subsection{Malware Classification and Speculative Defense} 

\noindent
Early feasibility studies demonstrated that hardware performance counters could successfully identify anomalous execution patterns without incurring the overhead of software instrumentation. Tang \textit{et al.}~\cite{tang2014unsupervised} pioneered this approach by using unsupervised machine learning to classify general malware signatures. Following the disclosure of transient execution vulnerabilities, researchers adapted these techniques specifically for Spectre and Meltdown. Li \textit{et al.}~\cite{li2019detecting} provided foundational proof-of-concept thresholds, demonstrating that hardware vulnerabilities leave distinct footprints in cache and branching metrics. Kotha \textit{et al.}~\cite{kotha2025datadriven} expanded on this by proving the feasibility of high-accuracy Multi-Layer Perceptrons (MLPs) for detecting unthrottled speculative bursts. While these early works successfully proved that ML-based detection is possible, their primary weakness lies in their evaluation under isolated, ideal conditions, failing to account for the signal degradation caused by real-world system noise.

To address the limitations of single-device evaluations, subsequent research began exploring cross-platform telemetry. Larson \textit{et al.}~\cite{larson2021realtime} conducted basic cross-platform measurements for edge devices, establishing that HPC behavior diverges across architectures. However, their work primarily focused on general malicious attacks and did not map the specific evasion space of transient execution vulnerabilities or evaluate adversarial adaptation. 

\subsection{The SoK Era: Recognizing Fragility}

\noindent
Recognizing that sophisticated attackers will attempt to evade static thresholds, Li \textit{et al.}~\cite{li2022detecting} introduced the concept of ``Adversarial Pacing,'' proving that attackers can bypass detection by artificially limiting their execution throughput. While their work effectively modeled the attacker's evasion strategy, it was limited to a single architectural domain. More recently, the Systematization of Knowledge (SoK) by Kosasih \textit{et al.}~\cite{kosasih2024sok} comprehensively analyzed 50 prior works, concluding that the vast majority of proposed HPC detectors remain critically vulnerable. The SoK identified a systemic failure in the literature to account for ``Domain Shifts''---specifically highlighting the lack of realistic threat models that combine architectural variance, environmental noise, and adversarial pacing.

\subsection{Research Gap and Contributions of this Work}

\noindent
As summarized in Table~\ref{tab:novelty_comparison}, this work directly answers the SoK's critique by addressing the fundamental research gap: the absence of a multidimensional threat model evaluation. We provide the first unified empirical characterization that simultaneously maps cross-ISA drift, multi-variant signatures, system noise interference, and adversarial pacing. Furthermore, we evaluate the ``quality'' of this characterization by quantifying the detection collapse of standard models (Section~\ref{sec:implications}), providing the realistic threat model evaluation that the current state-of-the-art lacks.

\begin{table}[!ht]
\caption{Comparison of the Characterization Matrix Against Prior HPC Security Literature. CI = Cross-ISA, MV = Multi-Variant, SN = System Noise, AP = Adversarial Pacing.}
\label{tab:novelty_comparison}
\begin{center}
\resizebox{\columnwidth}{!}{%
\begin{tabular}{|l|c|c|c|c|p{3.5cm}|}
\hline
\textbf{Reference} & \textbf{CI} & \textbf{MV} & \textbf{SN} & \textbf{AP} & \textbf{Primary Objective} \\
\hline
\hline
Tang \textit{et al.}~\cite{tang2014unsupervised} & $\times$ & $\times$ & $\times$ & $\times$ & General anomaly classification \\
\hline
Li \textit{et al.}~\cite{li2019detecting} & $\times$ & $\times$ & $\times$ & $\times$ & Proof-of-concept ML thresholds \\
\hline
Larson \textit{et al.}~\cite{larson2021realtime} & \checkmark & $\times$ & $\times$ & $\times$ & Basic cross-platform measurement \\
\hline
Kotha \textit{et al.}~\cite{kotha2025datadriven} & $\times$ & $\times$ & \checkmark & $\times$ & High-accuracy MLP feasibility \\
\hline
Kosasih \textit{et al.}~\cite{kosasih2024sok} & $\times$ & \checkmark & $\times$ & $\times$ & Systematization of ML failures \\
\hline
Li \textit{et al.}~\cite{li2022detecting} & $\times$ & $\times$ & \checkmark & \checkmark & Attacker throughput vs. detection \\
\hline
\textbf{This Work} & \textbf{\checkmark} & \textbf{\checkmark} & \textbf{\checkmark} & \textbf{\checkmark} & \textbf{Empirical characterization of multi-dimensional domain shifts} \\
\hline
\end{tabular}%
}
\end{center}
\end{table}
\section{Experimental Methodology}
\label{sec:methodology}
%

\noindent
To capture the microarchitectural variance of transient execution attacks, our experimental framework systematically isolates the variables that distort HPC telemetry. We evaluate three distinct attack profiles (Spectre V1 with Flush+Reload, Spectre V2 with a Flush+Reload-like timing leak, and Spectre V1 with Prime+Probe) across a multidimensional matrix of hardware topologies, environmental interference, and adversarial traffic shaping. 

\subsection{Hardware Testbed and Architectural Diversity}
\noindent
We deploy our attack implementations across four distinct microarchitectures, representing a spectrum of speculative depths, cache replacement policies, and instruction sets:

\begin{itemize}
    \item \textbf{Intel Core i7-3537U (Ivy Bridge):} Serves as our primary baseline representing deep-speculation pipelines with strictly inclusive Last-Level Caches (LLC). \textit{(Referred to hereafter and in figures as Intel 1).}
    \item \textbf{Intel Core i5-7200U (Kaby Lake):} Acts as a secondary inclusive-cache baseline to isolate and quantify intra-ISA (generational) telemetry variance. \textit{(Referred to hereafter and in figures as Intel 2).}
    \item \textbf{ARM Cortex-A76:} Represents modern RISC architectures featuring a \textit{non-inclusive} L3 victim cache, enabling the study of cross-ISA telemetry shifts in decoupled hierarchies.
    \item \textbf{AMD A4-5000 (Jaguar):} Represents embedded x86-64 microarchitectures featuring a strictly banked, inclusive L2 cache topology, providing the foundation for our hardware-intrinsic mitigation analysis.
\end{itemize}

\noindent
The exact hardware specifications, cache topologies, and software environments for all evaluated platforms are summarized in Table~\ref{tab:testbed_specs}.

\begin{table}[htbp!]
\caption{Experimental Testbed Specifications}
\label{tab:testbed_specs}
\begin{center}
\begin{tabular}{|>{\raggedright\arraybackslash}p{1.2cm}|>{\raggedright\arraybackslash}p{1.5cm}|>{\raggedright\arraybackslash}p{1cm}|>{\raggedright\arraybackslash}p{1.1cm}|>{\raggedright\arraybackslash}p{1.5cm}|}
\hline
\textbf{Processor} & \textbf{Micro architecture}& \textbf{Cores / Threads} & \textbf{Cache (L1d / L2 / L3)} & \textbf{OS \& Kernel} \\
\hline
\hline
Intel Core i7-3537U & Ivy Bridge (22nm) & 2C / 4T & 32KB / 256KB / 4MB (Inclusive) & Kali 2025.3 (Linux 6.12) \\
\hline
Intel Core i5-7200U & Kaby Lake (14nm) & 2C / 4T & 32KB / 256KB / 3MB (Inclusive) & Kali 2025.3 (Linux 6.12) \\
\hline
ARM Cortex-A76 & ARM v8.2-A (7nm) & 4C / 4T & 64KB / 512KB / 2MB (Non-Inclusive) & Debian 12 (Linux 6.1) \\
\hline
AMD A4-5000 & Jaguar (28nm) & 4C / 4T & 32KB / 2MB (Inclusive [4-Bank]) / None & Kali 2025.3 (Linux 6.16) \\
\hline
\end{tabular}%
\end{center}
\end{table}

\noindent \textbf{Hardware Controls and Scalability.} We deliberately chose these platforms as controlled baselines to isolate unfiltered hardware behavior. Newer processors contain hidden microcode patches (such as branch history flushing) that mask raw hardware signals. To accurately map the true, unmitigated footprint of the attacks, we needed systems completely free of these black-box filters. Most importantly, the primary architectural difference we evaluated—inclusive versus non-inclusive cache design—remains highly relevant today. Modern enterprise servers (such as Intel Xeon Scalable and AMD Zen) have shifted toward non-inclusive or exclusive caches to manage massive memory capacities. Therefore, the hardware dynamics we isolated on our test machines apply directly to today's high-performance processors.

\subsection{Adversarial Pacing and Traffic Shaping}
\noindent
To model a sophisticated adversary attempting to blend into benign background noise, we introduce traffic shaping. We deliberately mix four pacing modes to evaluate attack robustness and throughput-confidence trade-offs by modulating the frequency and duration of speculative cache-probing operations:

\begin{itemize}
    \item \textbf{Unthrottled Mode:} Performs continuous speculative memory reads and subsequent cache-line reloads with no explicit throughput target. This serves as the baseline control condition, representing the maximum microarchitectural footprint achievable by the exploit.
    
    \item \textbf{Constant Rate Mode:} Sets a global target for successful data leakage in characters per second. It normalizes throughput and reduces timing drift by inserting a calculated microsecond-level delay immediately after each secret-access-and-reload cycle to warp the resulting HPC distribution.
    
    \item \textbf{Burst-All Mode:} Executes a rapid, back-to-back sequence of speculative reads until the target payload is reached, followed by an extended sleep period so the total activity aligns to a one-second window. This stresses the stability of telemetry under clustered, high-intensity activity followed by microarchitectural silence.
    
    \item \textbf{Batch-Rate Mode:} Organizes speculative probes into discrete, fixed-size batches. The system enforces a per-batch timing target proportional to the batch size divided by the rate, modeling a highly structured, periodic attack traffic designed to mimic the temporal patterns of legitimate background processes.
\end{itemize}

\subsection{The Interference Matrix and Benign Baselines}
To neutralize scheduling artifacts, our background load engine leverages NumPy and SciPy. These primitives explicitly release the CPython Global Interpreter Lock (GIL), allowing background tasks to execute as unthrottled concurrent threads across separate physical cores to induce four distinct architectural states:

\begin{itemize}
    \item \textbf{Idle:} The baseline condition running with no active interference threads, serving as the optimal, quiet environment for the attacker.
    \item \textbf{CPU-Intensive:} A continuous, multi-threaded workload that computes dot products and calculates eigenvalues on random floating-point matrices (base size of $200 \times 200$). This introduces severe arithmetic and scheduling contention.
    \item \textbf{Memory-Pressure:} An array-manipulation workload that continuously allocates, sorts, and computes Fast Fourier Transforms (FFTs) on large floating-point arrays ($10^6$ elements). The generator maintains a rolling buffer of the 10 most recent arrays to force continuous cache-line replacement and thrash the LLC.
    \item \textbf{Mixed Workload:} A highly variable workload designed to model realistic, multi-faceted behavior. It rapidly cycles through pseudo-inverse matrix calculations, normal distribution statistics (mean and standard deviation), and 1D FFT signal processing.
\end{itemize}

\noindent
\textbf{Benign Baseline Workloads:} Separately from the background interference, when establishing the non-malicious "Benign" execution baselines (evaluated in Section 5), we avoid using simple idle sleep states. Instead, the framework executes a diverse suite of standard system utilities to model legitimate application telemetry. These workloads include \texttt{stress-ng} for targeted CPU and virtual memory pressure, \texttt{openssl sha256} for cryptographic hashing, \texttt{gcc} for source compilation, and \texttt{tar}/\texttt{dd} for heavy file I/O operations.
\subsection{Automated Data Collection and Alignment}
\noindent
To maintain reproducibility and strict temporal alignment, we orchestrate the data collection using a custom \texttt{batch\_auto} pipeline. 

For each attack bucket, the matrix dictates 40 total runs (10 attack runs across each of the 4 primary background modes). To maintain strict run isolation, the pipeline enforces the following sequence: the chosen interference condition is initiated prior to attack execution, the attack executes and logs timestamped metadata, the interference is terminated, and a stabilization gap is enforced before the next run. This rigid coupling assures that the varying combinations of attack families, pacing models, and interference conditions remain perfectly aligned with the sampled HPC telemetry, enabling highly accurate comparative evaluations.
\section{Empirical Characterization of Multi-Dimensional Domain Shifts}
\label{sec:characterization}

\noindent
To establish a clear baseline for domain adaptation, we first analyze the ``ground truth'' HPC telemetry of transient execution attacks---defined here as the signatures generated by unthrottled exploits executing on an isolated, idle system. This serves as the idealized reference point for characterizing the transition from optimal signal visibility to the ``signal collapse'' observed under heavy system noise and intentional traffic shaping. By decoupling these layers, we isolate the specific variance induced by the hardware architecture, the leakage mechanism, and the adversary's defensive evasion strategy.

\subsection{Cross Device Telemetry: The Impact of Microarchitecture}
\label{sec:baseline_device}

%

\noindent
A core fallacy in portable hardware security is the assumption that identical attack software generates uniform telemetry among different processors. To quantify this cross-device variance, we evaluate the unthrottled Spectre V1 Flush+Reload implementation against a benign idle baseline across our primary hardware profiles: the Intel Core i7-3537U, ARM Cortex-A76, and AMD Jaguar. 

To conduct a fair cross-ISA comparison, we abstract the telemetry into normalized ratios rather than raw event counts. Relying on raw counts introduces severe bias due to variations in clock frequencies and pipeline widths across architectures. Instead, we monitor two generalized architectural metrics: the Cache Miss Ratio (calculated as Last Level Cache misses divided by Last Level Cache references) and the Branch Miss Ratio (calculated as global branch mispredictions divided by global branch references).

\begin{figure}[ht]
    \centering
    \includegraphics[width=1.0\linewidth]{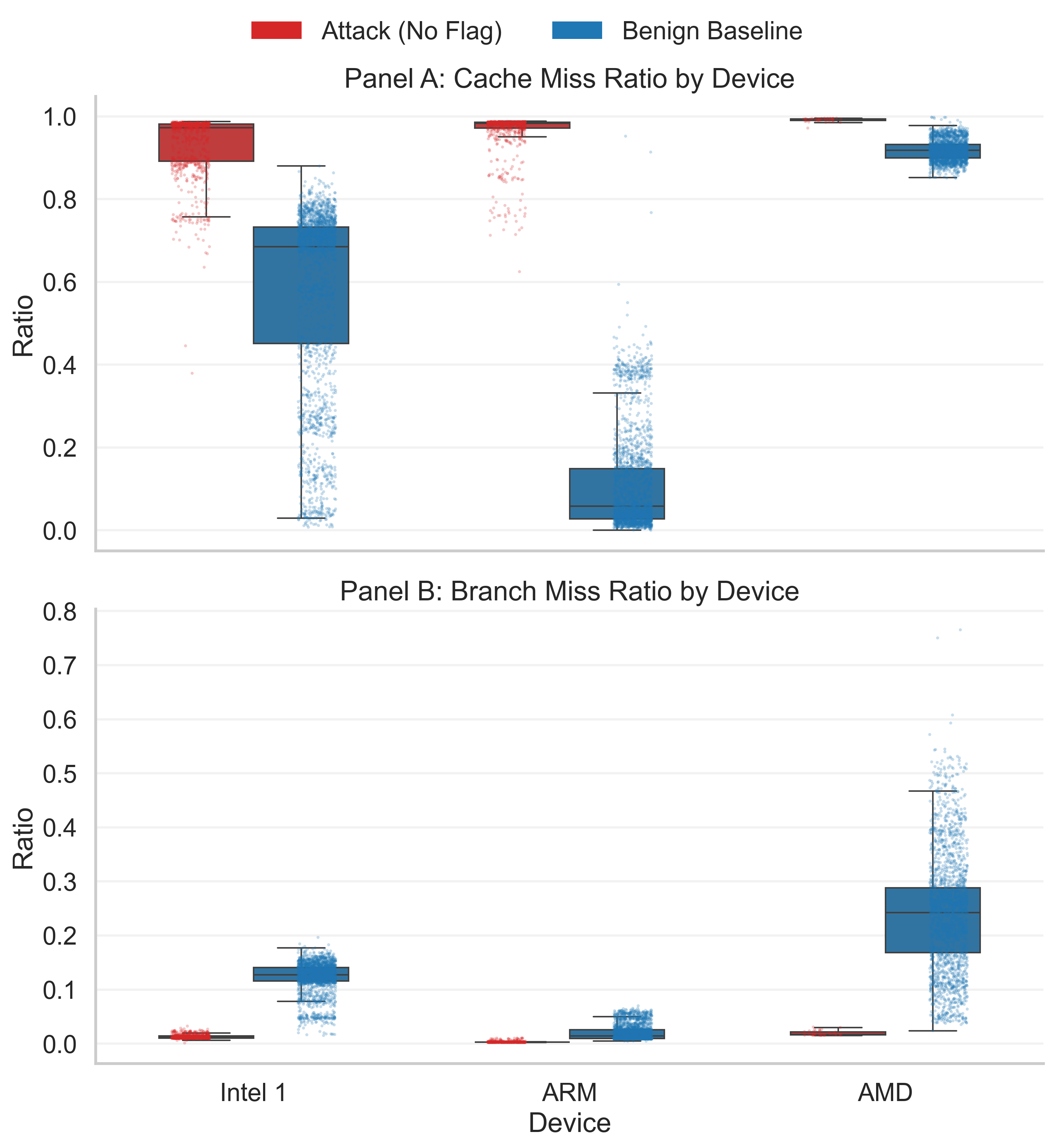}
    \caption{Baseline HPC telemetry comparing unthrottled Spectre V1 (red) against benign idle execution (blue). Panel A illustrates the Cache Miss Ratio, while Panel B illustrates the Branch Miss Ratio. The overlaid scatter plots reveal severe microarchitectural variance in the baseline system noise.}
    \label{fig:cross_device_v1}
\end{figure}

As visualized in Figure \ref{fig:cross_device_v1}, the distributions of both the attack and the benign baseline differ drastically across the microarchitectures.

In Panel A, the attack consistently forces a Cache Miss Ratio near 1.0 across all devices, which is expected given the aggressive cache flushing inherent to the Flush+Reload channel. However, the benign baseline reveals massive architectural divergence. On the Intel Core i7 platform, the benign cache miss ratio spans a wide variance envelope, creating a measurable gap between normal execution and the attack. 

Conversely, the AMD Jaguar exhibits a naturally high and tightly clustered benign Cache Miss Ratio (Median: $0.92$, IQR: $0.90\text{--}0.93$). Compared to the Intel platform's benign median of $0.69$, this native architectural behavior drastically compresses the separability margin between benign and malicious execution, mathematically proving the risk of cross-ISA false positives.

Panel B demonstrates a similar divergence in the Branch Predictor. Across all platforms, the attack execution creates a highly predictable control flow, resulting in a low, tightly clustered Branch Miss Ratio. However, the benign system noise varies wildly. The ARM Cortex-A76 maintains a low benign branch miss ratio, whereas the AMD Jaguar exhibits massive variance with an interquartile range stretching far above the other platforms.

This baseline data proves that the separation boundaries between benign and malicious execution are strictly hardware-dependent. A static machine learning classifier trained to detect the wide cache miss separation on an Intel processor will completely fail to generalize to the compressed, noisy environment of the AMD Jaguar.           
\subsection{Cross-Variant Fingerprints: Mechanisms and Channels}
\label{sec:baseline_variant}

\noindent
While Section \ref{sec:baseline_device} establishes the baseline variance across architectures, a secondary domain shift occurs when an attacker alters their exploit strategy on a specific, fixed hardware target. To quantify this intra-ISA variance, we evaluate three distinct attack variants on a single platform—the Intel Core i5-7200U—under Idle conditions: Spectre V1 via Flush+Reload (V1\_fr), Spectre V2 via Flush+Reload (V2\_fr), and Spectre V1 via Prime+Probe (V1\_pp). 

\begin{figure}[ht]
    \centering
    \includegraphics[width=1.0\linewidth]{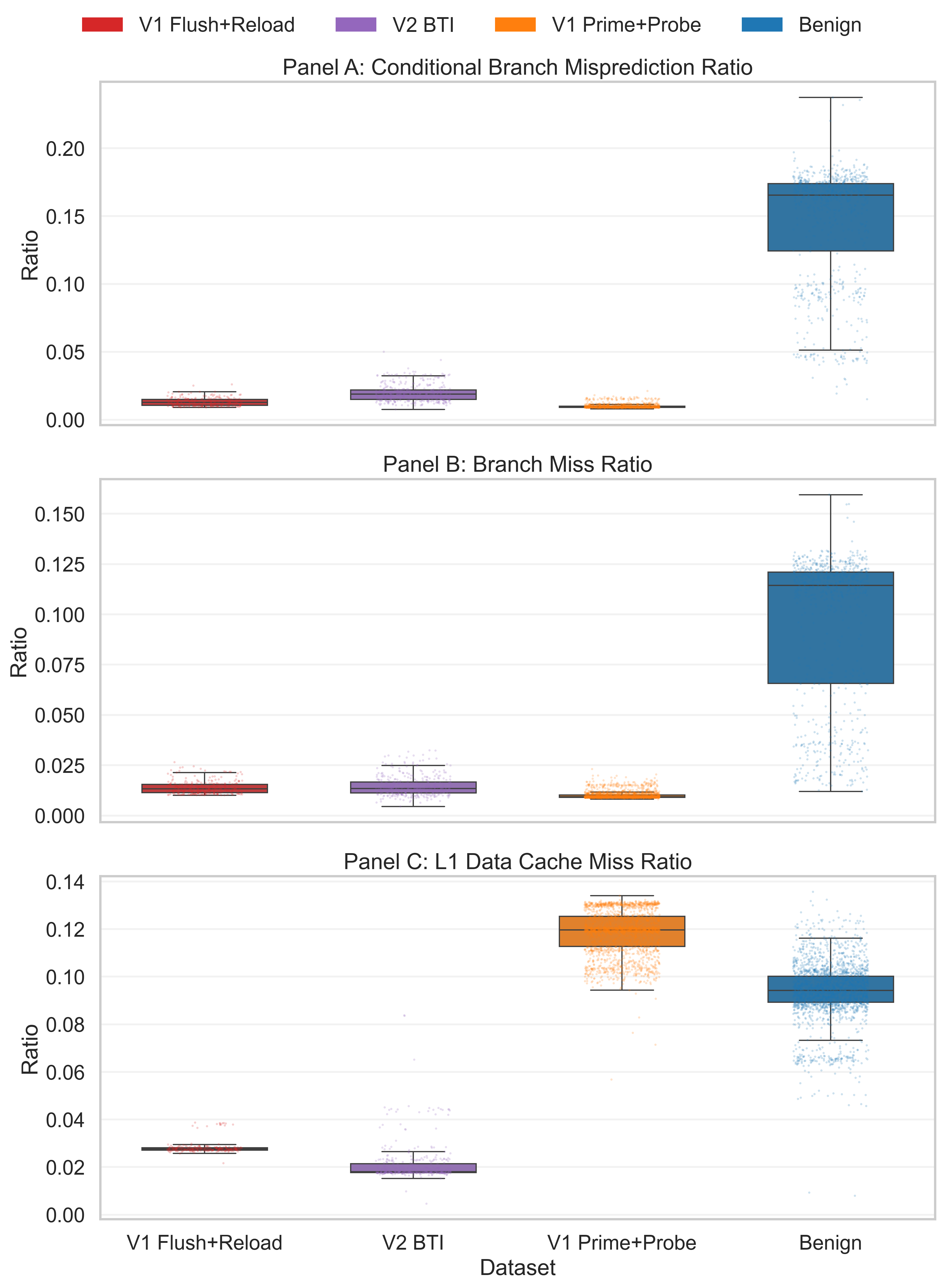}
    \caption{Baseline HPC fingerprints for attack variants on the Intel Core i5-7200U platform. Panels A and B highlight the statistical similarity in speculative stability across different branch metrics, while Panel C illustrates the L1 Data Cache Miss Ratio.}
    \label{fig:cross_variant}
\end{figure}

This analysis models a realistic adversary who, rather than dealing with different platforms, specializes their exploit for a specific victim machine but alternates between different microarchitectural triggers or leakage channels to bypass signature-based monitors. Because these variants manipulate different components of the same microarchitecture, they generate highly specific telemetry fingerprints. Figure \ref{fig:cross_variant} illustrates this divergence across three primary microarchitectural vectors. To ensure robustness against temporal aliasing in the hardware counters, we utilize self-contained retirement and misprediction ratios from within our synchronized collection batches.

Panels A and B isolate the impact of the speculation triggers via two distinct lenses: the Conditional Branch Misprediction Ratio and the Global Branch Miss Ratio. Whether utilizing a conditional bounds check bypass (V1\_fr, V1\_pp) or indirect branch target injection (V2\_fr), all three attack variants show remarkably similar fingerprints characterized by low-magnitude, tightly clustered ratios. 

In the branching domain (Panel A), Spectre V2 (V2\_fr) exhibits a slightly higher Conditional Branch Misprediction Ratio (Median: $0.018$) compared to the Spectre V1 variants (Median: $0.012$), reflecting the additional overhead of indirect branch poisoning. However, this variation remains statistically insignificant when compared to the massive variance and elevated median of the naturally noisy benign baseline (Median: $0.165$). This consistency occurs because unthrottled attack loops successfully train the underlying predictors to favor the speculative path, resulting in a stable, low-entropy signature. In contrast, the benign baseline represents diverse architectural paths and stochastic system activity, leading to significantly higher instability that overshadows the subtle differences between exploit mechanisms.

Panel C highlights the divergence in leakage channels. While V1\_fr and V1\_pp share a similar speculation trigger, Prime+Probe (V1\_pp) infers data by measuring set-based disturbance across the cache hierarchy. This necessitates constructing eviction sets and thrashing the L1 cache, pushing its median L1 Data Cache Miss Ratio to $0.119$. In contrast, the surgical Flush+Reload variants create a low-impact footprint (Median: $0.027$), placing Prime+Probe's severe cache contention visibly above both the F+R variants and the benign baseline (Median: $0.094$). 

These distinct fingerprints prove that the variance envelope is mechanism-dependent even on a constant hardware target. A detector optimized for the cache signatures of Flush+Reload will likely fail to recognize the heavy set-based contention of Prime+Probe, offering a clear evasion path for sophisticated adversaries.      
\subsection{Signal Degradation Under Environmental Noise}
\label{sec:degradation}
%

\begin{figure}[ht]
    \centering
    \includegraphics[width=1.0\linewidth]{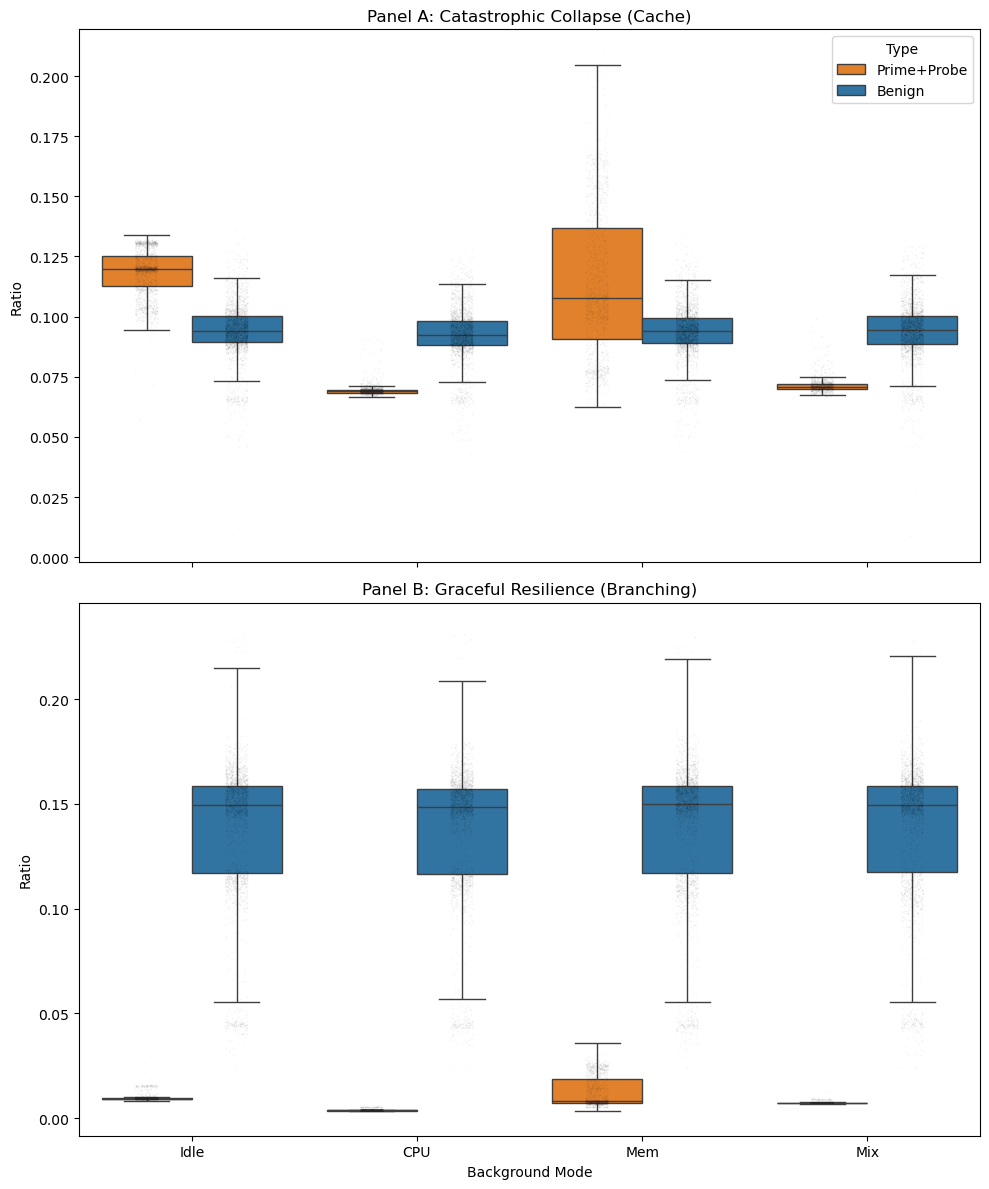}
    \caption{Side-by-side comparison of telemetry degradation. Panel A illustrates the catastrophic collapse of the L1 Cache Miss Ratio due to resource contention and signal inversion. Panel B demonstrates the graceful resilience of the Conditional Branch Misprediction Ratio, which maintains a clear margin of separability across all noise modes.}
    \label{fig:noise_stress}
\end{figure}

\noindent
While the isolated fingerprints in Section 4.2 provide clear microarchitectural benchmarks, real-world deployment occurs within an unpredictable environment where background operations compete for limited hardware resources. To quantify the stability of these signals, we subject the V1 Prime+Probe attack and a Benign baseline to a multi-dimensional interference matrix on the Intel Core i5-7200U platform. Figure~\ref{fig:noise_stress} illustrates the divergence between metrics that degrade gracefully and those that collapse entirely under system load.

The data reveals two opposing microarchitectural behaviors. In Panel A, the L1 Data Cache Miss Ratio---a primary metric for cache-based side channels---exhibits a catastrophic loss of separability. Under resource-contention-heavy workloads ({\it e.g.,} CPU-bound and memory-intensive background tasks), we observe that the ``signal inversion'' is not simply a magnitude shift but a frequency-domain overlap between benign cache thrashing and malicious eviction. 

This suggests that ALU-heavy contention actively throttles the attacker's execution frequency, ``silencing'' the cache-miss signature. Furthermore, under memory pressure (Mem), the benign L1 cache distribution expands significantly (Median: $0.094$, Max: $0.134$) to completely swallow the Prime+Probe attack's median signature (Median: $0.108$). Because the variance of the benign noise overlaps the attack footprint, static L1 cache thresholds are rendered mathematically ineffective in this domain.

In contrast, Panel B demonstrates that speculative branching metrics degrade gracefully. Even under heavy memory contention, the Conditional Branch Misprediction Ratio of the attack footprint (Median: $0.008$, Max: $0.036$) maintains a clear, near-constant margin of separability from the naturally noisy benign baseline (Median: $0.150$). This resilience stems from the dedicated nature of the Branch Prediction Unit (BPU); while the cache is a shared global resource easily thrashed by background tasks, the BPU remains highly sensitive to the repetitive, tight-loop logic characteristic of Spectre exploits. 

Even as the system reaches peak interference in the Mixed mode, the attack's ``low-entropy'' branching signature remains distinct. This comparison illustrates a fundamental bottleneck for hardware-based malware detection: the selection of features must be context-aware. Relying on cache-level telemetry alone invites evasion through environmental noise, whereas integrating resilient branching metrics offers a stable foundation for domain-adaptive models.        
\subsection{Adversarial Pacing \& Traffic Shaping}
\label{sec:pacing}
%

\noindent
While previous sections analyze the impact of extrinsic environmental noise, this section assesses the effectiveness of intentional, intrinsic adversarial manipulation. A sophisticated attacker can trade execution throughput for microarchitectural stealth by using traffic shaping---modulating the frequency and duration of speculative bursts to distort the resulting Hardware Performance Counter (HPC) distributions. 

\begin{figure}[ht!]
    \centering
    \includegraphics[width=1.0\linewidth]{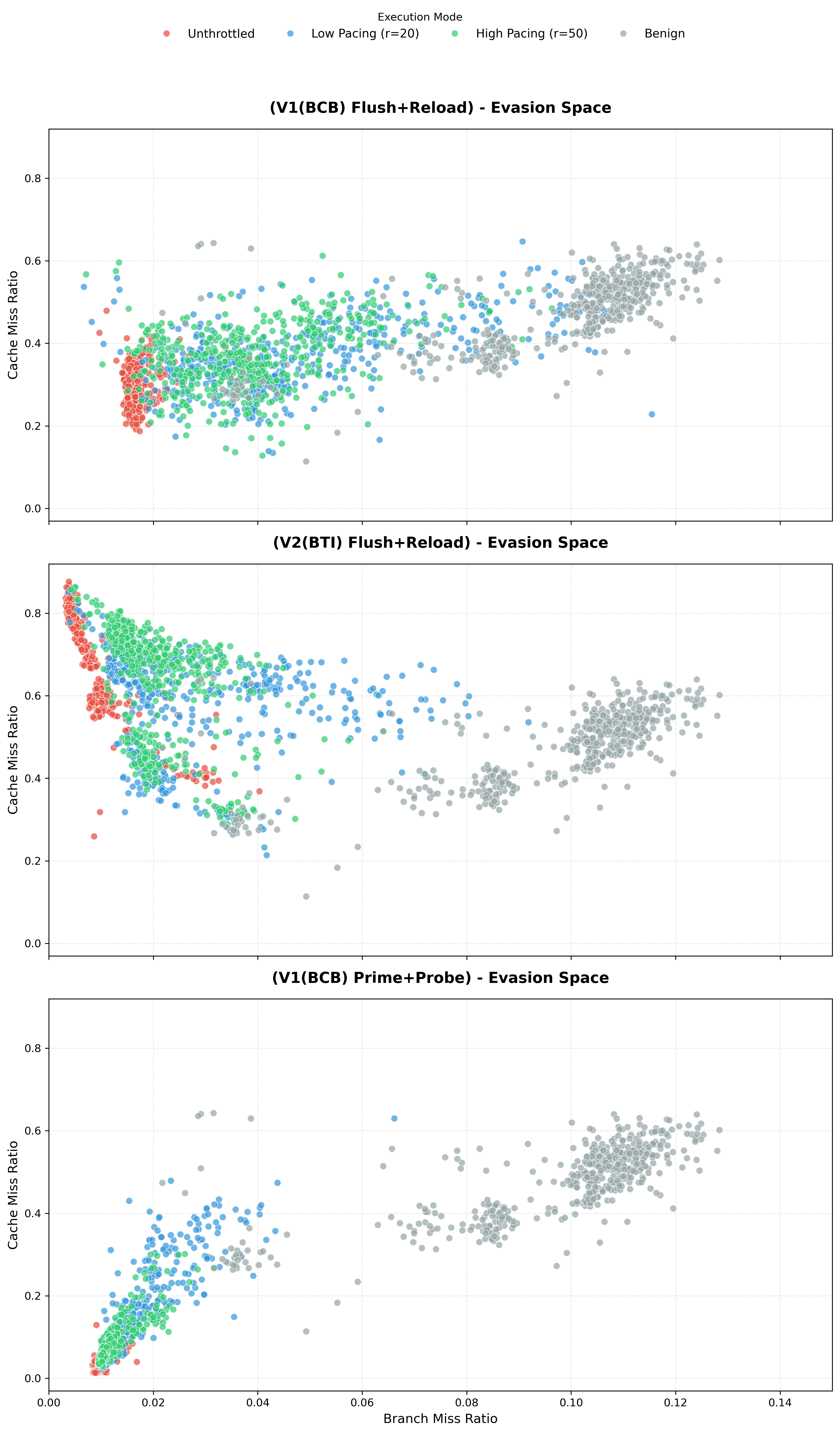}
    \caption{Evasion Space trajectories for Spectre V1 and V2. Flush+Reload variants (top/middle) maintain high cache-intensity signatures, whereas V2(BTI) resists branching entropy diffusion. Conversely, Prime+Probe (bottom) exhibits a diagonal signal collapse, merging indistinguishably into the benign baseline under heavy pacing.}
    \label{fig:evasion_comparison}
\end{figure}

To quantify this ``Throughput-Visibility Trade-off,'' we evaluate three distinct attack primitives: V1(BCB) Flush+Reload, V2(BTI) Flush+Reload, and V1(BCB) Prime+Probe. As illustrated in Figure \ref{fig:evasion_comparison}, we map the resulting two-dimensional evasion space using global branch and cache miss ratios. By subsampling the telemetry to 500 random points per execution mode, we isolate the specific microarchitectural ``trajectories'' that occur as an adversary transitions from an unthrottled state to high-pacing regimes ($r=50$).

The empirical data reveals three distinct evasion behaviors that are highly dependent on the underlying side-channel mechanism and the speculative trigger:

\begin{itemize}
    \item \textbf{V1(BCB) Flush+Reload (Horizontal Entropy Diffusion):} In the top panel, the attack exhibits a primarily horizontal migration. As pacing increases, the signature maintains a high Cache Miss Ratio (consistently between 0.2 and 0.5) but slides rightward along the X-axis. This indicates that while temporal shaping reduces the frequency of cache evictions, it simultaneously increases the microarchitectural entropy within the branch predictor. This diffusion pushes the signature into the high-variance region of the benign clusters, attempting to facilitate evasion through signal blurring.
    
    \item \textbf{V2(BTI) Flush+Reload (Top-Left Precision Retention):} The middle panel demonstrates a ``top-left retention'' strategy. Unlike V1, this variant remains pinned in the high-cache/low-branch-miss zone even under heavy pacing ($r=50$). This indicates that the Branch Target Injection (BTI) mechanism is fundamentally more surgical; the adversary maintains near-perfect branch prediction accuracy while executing high-frequency flushes. This creates a ``loud'' microarchitectural outlier that resists blending into the stochastic noise of legitimate applications.
    
    \item \textbf{V1(BCB) Prime+Probe (Diagonal Signal Collapse):} The bottom panel showcases the most elusive primitive. It follows a tight diagonal collapse trajectory where both cache intensity and branching signatures degrade proportionally. In the high-pacing regime, the attack becomes functionally indistinguishable from the lowest-intensity benign samples. Unlike Flush+Reload variants, which are anchored by high-intensity cache footprints, paced Prime+Probe achieves near-total microarchitectural invisibility.
\end{itemize}

This divergence underscores a fundamental challenge for unified hardware detectors. While global metrics remain sensitive to ``noisy'' primitives like Flush+Reload, they are highly susceptible to evasion by ``quiet'' primitives like paced Prime+Probe. Our results highlight the critical necessity for detectors to move beyond static, global thresholds and instead adopt variant-aware models that can identify ``low-entropy'' speculative states even when the primary cache-level signatures have been successfully shaped into benign magnitudes.             
\subsection{Implications for Adaptive Detection}
\label{sec:implications}
%

\noindent
The empirical variance characterized across Subsections~\ref{sec:baseline_device} through \ref{sec:pacing} proves that Spectre microarchitectural signatures are volatile functions of the hardware platform, exploit mechanism, and environmental context, yielding three critical consequences for resilient monitor design.

\subsubsection{Hardware Domain Shift and Portability}
The cross-device divergence characterized in Section~\ref{sec:baseline_device} proves raw telemetry is non-portable due to the frequency and pipeline biases described earlier~\cite{kotha2025datadriven,li2022detecting}. Because localized hardware traits—such as the compressed separability margin caused by the AMD Jaguar's naturally elevated cache noise floor—drastically inflate false-positive rates, direct cross-platform deployments suffer from severe signature degradation~\cite{kosasih2024sok}. This ``Hardware Domain Shift'' underscores the necessity for architecture-aware normalization or transfer learning to dynamically calibrate decision thresholds to each device's native noise floor.

\subsubsection{Feature Stability vs. Mechanism Variance}
The intra-ISA variance characterized in Section~\ref{sec:baseline_variant} highlights the danger of mechanism-specific overfitting. An attacker can retain the same speculative trigger while entirely altering the leakage channel's footprint (Flush+Reload vs. Prime+Probe). Attention-based feature importance reveals that LLC-related metrics (LLC Misses and LLC Load Misses) serve as the most stable indicators across platforms, significantly outperforming volatile L1 cache features easily thrashed by background system noise. Robust detection must therefore prioritize resilient branching ratios and stable LLC telemetry.

\subsubsection{The Necessity of Domain Adaptation}
The synthesis of environmental noise and traffic shaping formalizes the requirement for Domain Adaptation (DA). 
To evaluate our characterization data, we execute a zero-shot domain transfer study using a baseline LSTM network (64 hidden units, 32 dense units, sigmoid activation) trained on the Intel 1 (Idle) dataset ($80/20$ split, Adam optimizer, binary cross-entropy, batch size 64, early stopping). To cleanly isolate instant-state thresholds from temporal smoothing dependencies, the model configures its input layer to process a single timestep (sequence length = 1, 4 features).

This single-timestep configuration acts as an analytical control. Standard sliding-window sequence models suffer from acute temporal overfitting to the source domain's instruction strides. Under adversarial traffic shaping (e.g., Constant-Rate or Batch-Rate modes), the adversary systematically shifts speculative access intervals, shattering temporal dependencies and causing severe sequence misalignment. Removing the temporal window cleanly tests raw threshold resilience under active pacing.

As summarized in Table~\ref{tab:transfer_results}, while the model achieves near-perfect separability on its source domain ($F1 \approx 1.00$) and generalizes across intra-ISA generational drift (Intel 2, $F1 \approx 0.99$), performance degrades under memory pressure on the same host machine (Intel 1 Memory Pressure, $F1 = 0.54$). Most critically, simultaneous cross-ISA transfer and adversarial pacing induce complete detection collapse ($F1 = 0.00$, $Accuracy = 0.50$), performing no better than random chance due to the lack of tracking over shaped telemetry intervals. This failure mathematically proves that static, single-input signatures established in isolated environments are blind to domain-shifted telemetry, meaning future detectors must transition to adaptive monitoring architectures.

\begin{table}[ht!]
\caption{Zero-Shot Transfer Performance: Baseline LSTM trained on Intel 1 (Idle), tested against characterized domain shifts.}
\label{tab:transfer_results}
\begin{center}
\begin{tabular}{|>{\raggedright\arraybackslash}p{2.3cm}|c|c|>{\raggedright\arraybackslash}p{2.5cm}|}
\hline
\textbf{Target Domain (Test Set)} & \textbf{F1-Score} & \textbf{Accuracy} & \textbf{Observed Effect} \\
\hline
\hline
Intel 1 (Idle) & 1.00 & 1.00 & Baseline (Source Data) \\
\hline
Intel 2 (Idle) & 0.99 & 0.99 & Cross-Generation Drift \\
\hline
Intel 1 (Memory Pressure) & 0.54 & 0.69 & Environmental Noise \\
\hline
ARM (Paced Attack) & 0.00 & 0.50 & \textbf{Cross-ISA Signal Collapse} \\
\hline
\end{tabular}
\end{center}
\end{table}

\noindent 
The AMD Jaguar platform is omitted from Table~\ref{tab:transfer_results} because its functional Flush+Reload metrics match the ARM baseline's transfer degradation. Crucially, for set-based interference, the Jaguar microarchitecture presents a unique hardware anomaly. As detailed in Section \ref{sec:bottlenecks}, the Jaguar's banked L2 cache hierarchy and aggressive L2 prefetching physically neutralize the Prime+Probe channel. Rather than repeating redundant ML failure rates, the next section provides a microarchitectural post-mortem of how intrinsic hardware constraints can physically dismantle specific attack channels before their signatures can reach the telemetry layer.       
\section{Microarchitectural Bottlenecks: A Case Study of Prime+Probe}
\label{sec:bottlenecks}
%

\noindent
While Prime+Probe is often considered an adaptable side-channel due to its independence from shared memory, empirical results show severe non-portability across disparate cache hierarchies~\cite{li2022detecting}. As demonstrated in Section~\ref{sec:baseline_device}, Prime+Probe signatures exhibit near-perfect separability on Intel and ARM platforms but collapse completely into indistinguishability on the AMD Jaguar. This section presents a microarchitectural post-mortem proving this failure stems from physical memory hierarchy constraints rather than timing measurement limitations.

\subsection{Beyond Timing: Validating the Channel Resolution}
\noindent
Failed cache exploits are frequently misattributed to insufficient timer resolution ({\it e.g.,} low-frequency \texttt{rdtsc} execution). However, our timing analysis confirms that the Jaguar platform provides sufficient granularity to cleanly differentiate L1 hits from L2/DRAM misses. While the ``Probe'' phase successfully records latency deltas, these variations fail to correlate with malicious activity. This signature absence indicates that the attack fails to induce microarchitectural state changes, exposing an inherent state of \textit{Architectural Resistance} rather than measurement failure.

\subsection{The Inclusive L2 Paradox and Eviction Failure} \label{sec:l2_paradox}
\noindent
On Intel platforms, inclusive Last Level Caches (LLCs) enable reliable eviction from lower-level caches by saturating specific LLC sets, yielding a high-magnitude, low-variance telemetry footprint. Conversely, the Jaguar exposes an eviction misalignment paradox. While the L2 cache is nominally inclusive, its 2MB capacity is strictly divided into four physically routed banks. Because standard Prime+Probe loops lack the address-bit precision to evenly target these specific banks, the attacker populates the L2 unevenly. Failing to saturate the specific sets tracking the victim's data, the hardware is never triggered to enforce inclusive back-invalidation. Consequently, the loop remains microarchitecturally silent to the L1, generating a successful timer log of an unsuccessful physical eviction. This induces a complete \textit{signal washout}: while HPCs log L2 updates, the undisturbed L1 footprint prevents data leakage, rendering the L1 cache miss ratio statistically indistinguishable from benign idle noise.

\subsection{Hardware-Induced Telemetry Evasion via L2 Prefetching}
\noindent
The AMD Jaguar’s L2 hardware prefetcher introduces a concurrent mechanism for telemetry dilution. Triggered by high-frequency Prime+Probe loop access intervals, the prefetcher aggressively streams lines into the L2 before the ``Probe'' phase registers a timing delta, effectively erasing the channel's temporal footprint. We must demarcate these functional boundaries: the L2 banking topology mismatch (Section \ref{sec:l2_paradox}) is the primary structural bottleneck causing physical exploit failure. Conversely, the prefetcher acts as the primary telemetry masking agent. This is empirically validated by toggling MSR \texttt{0xC0011022}, where disabling the prefetcher recovers the latency signal delta by only 12\%, leaving the remaining 88\% bound to the banked L2 layout mismatch. The prefetcher does not dismantle the exploit; rather, it visually dilutes the failed footprint, forcing telemetry to mirror benign idle execution.

\subsection{Summary: The Hard Limit of Portability}
\noindent
The Jaguar case study formalizes a fundamental constraint for hardware security: \textit{Observability is not microarchitecturally portable, even within a shared ISA.} The collapse of Prime+Probe signatures on the Jaguar proves that execution topology and directory structures dictate counter semantics independent of the instruction set. Consequently, hardware-based detectors cannot act as architecture-agnostic black boxes; a feature set optimized for the set-based contention of an inclusive-cache layout (Intel/ARM) remains fundamentally blind when deployed against a strictly banked L2 cache topology (AMD) sharing the same x86-64 instruction primitives and inclusive policy.
\section{Conclusion}
\label{sec:conclusion}
%

\noindent
This study provides a comprehensive examination of the microarchitectural observability of Spectre-class attacks under realistic deployment conditions. By analyzing the intersection of environmental noise, adversarial traffic shaping, and cross-device variance, this paper demonstrates that the ``ground truth'' signatures established in isolated laboratory environments are fundamentally fragile. 

Our results reveal a critical hierarchy of stealth. While Flush+Reload variants maintain a high-intensity cache footprint that is relatively resilient to pacing, Prime+Probe variants exhibit a catastrophic signal collapse, achieving near-total microarchitectural invisibility when shaped by an informed adversary.

Furthermore, the AMD Jaguar case study formalizes the existence of physical observability bottlenecks that remain ISA-independent. The failure of Prime+Probe due to L2 banking topology differences proves that hardware-based malware detection cannot rely on universal, static feature sets. These findings serve as a primary requirement for the development of adaptive, architecture-aware security monitors. 

To move beyond the limits of rigid thresholding, the field must prioritize domain-adaptation techniques capable of dynamically recalibrating decision thresholds. Based on this empirical foundation, our immediate future work will address these observability bottlenecks through advanced adaptive modelling. Specifically, we plan to explore hybrid architectures—integrating frameworks such as Domain-Adversarial Neural Networks (DANN) and XGBoost—to dynamically recalibrate decision thresholds in response to fluctuating noise floors and cross-ISA variance. Ultimately, the transition toward such context-aware, hybrid monitoring remains a microarchitectural necessity for securing modern speculative pipelines.

\section*{Acknowledgements}
\noindent
This work was supported in part by the National Science Foundation (NSF) under Grant CNS-2026675. Any opinions, findings, and conclusions or recommendations expressed in this material are those of the authors and do not necessarily reflect the views of NSF.

\paragraph{\textbf{Generative AI Disclosure}}
In accordance with IEEE policies, we disclose the use of large language models strictly for language refinement, structural organization, and LaTeX formatting. AI was not used to formulate hypotheses or generate data, and the authors take full responsibility for the paper's originality and scientific integrity.

\balance
\bibliographystyle{IEEEtran}
\bibliography{reference}

\end{document}